\documentclass[runningheads]{llncs}
\usepackage[utf8]{inputenc}
\usepackage{microtype}
\usepackage[breaklinks]{hyperref}
\usepackage{cleveref}
\usepackage{tikz}
\usepackage{listings}
\usepackage{zlmtt}
\usepackage{tabularx}
\usepackage{booktabs}
\usepackage{soul}
\usepackage{xcolor}
\usepackage{paralist}
\definecolor{orangeish}{HTML}{f1a340} 
\definecolor{grayish}{HTML}{f7f7f7} 
\definecolor{purpleish}{HTML}{998ec3} 

\definecolor{yellowish}{HTML}{DDAA33}
\definecolor{redish}{HTML}{BB5566}
\definecolor{blueish}{HTML}{004488}

\usepackage{enumitem}

\usepackage{nameref}

\makeatletter
\let\orgdescriptionlabel\descriptionlabel
\renewcommand*{\descriptionlabel}[1]{%
  \let\orglabel\label
  \let\label\@gobble
  \phantomsection
  \edef\@currentlabel{#1}%
  \let\label\orglabel
  \orgdescriptionlabel{#1}%
}
\makeatother

\usepackage{caption}
\usepackage{subcaption}

\newif\ifembargo
\embargofalse

\newif\ifverbose
\verbosefalse

\newcommand{\basicCodeStyle}{\ttfamily\small}
\newcommand{\keywordCodeStyle}{\bfseries}
\newcommand{\builtInCodeStyle}{\itshape\color{blueish}}
\newcommand{\typesCodeStyle}{\ttfamily\small\color{blueish}}

\lstdefinelanguage{TypeScript}{
  classoffset=0,
  morekeywords={typeof, new, true, false, catch, function, return, null, catch, switch, var, if, in, while, do, else, case, break, let, for, const, class, implements, enum, get, this},
  classoffset=1, 
  morekeywords={NetworkNode, string,EncryptedString,PaddedMsg,MessageType,DummyString},
  keywordstyle=\typesCodeStyle,
  classoffset=0,
  morecomment=[s]{/*}{*/},
  morecomment=[l]//,
  morestring=[b]",
  morestring=[b]'
}

\lstdefinelanguage{Noiselet}{
  classoffset=0,
  morekeywords={int, if, while, else},
  keywordstyle=\keywordCodeStyle,
  classoffset=1, 
  morekeywords={wait, gettime, last_kb_time,wait_t, send, load, store,sleep,reset,rnd,usleep},
  keywordstyle=\builtInCodeStyle,
  classoffset=0,
  morecomment=[s]{/*}{*/},
  morecomment=[l]//,
  morestring=[b]",
  morestring=[b]'
}

\crefname{lstlisting}{Listing}{Listings}

\newcommand{\ourprotocollong}{\textit{Deniable Instant Messaging} (DenIM)}
\newcommand{\ourprotocol}{DenIM}
\newcommand{\ourprotocolCaching}{DenIM Caching}
\newcommand{\ourprotocolEmbargo}{DenIM Caching+Embargo}
\newcommand{\normalmessage}{regular message}
\newcommand{\normalmessages}{regular messages}
\newcommand{\Normalmessages}{Regular messages}
\newcommand{\deniablemessage}{deniable message}
\newcommand{\deniablemessages}{deniable messages}
\newcommand{\dummymessage}{dummy message}
\newcommand{\dummymessages}{dummy messages}
\newcommand{\flagdrop}{\texttt{DUMMY}}
\newcommand{\flagnormal}{\texttt{REGULAR}}
\newcommand{\flagdeniable}{\texttt{DENIABLE}}
\newcommand{\flagblock}{\texttt{BLOCK\_REQUEST}}

\newcommand{\pdusize}{1134B}
\newcommand{\tlscipher}{TLS\_AES\_128\_GCM\_SHA256}
\newcommand{\rsaencryption}{RSAES-OAEP}

\newcommand{\attacker}{adversary}
\newcommand{\Attacker}{Adversary}
\newcommand{\attackers}{adversaries}

\newcommand{\indKeyExchange}{\textsc{Ind} 1}
\newcommand{\indSendMessage}{\textsc{Ind} 2}
\newcommand{\indKeyResponse}{\textsc{Ind} 3}
\newcommand{\indReceiveMessage}{\textsc{Ind} 4}
\newcommand{\indCoverTraffic}{\textsc{Ind} 5}

\newcommand\asmTrustedServer{A1}
\newcommand\asmTrustedDecoy{A2}
\newcommand\asmTrustedRcp{A3}
\newcommand{\asmDecoyNotRcp}{A4}
\newcommand\asmPatternContainment{A5}
\newcommand{\goalChanges}{DG1}
\newcommand{\goalPerformance}{DG2}
\ifverbose
\newcommand{\reqSend}{R1}
\newcommand{\reqReceive}{R2}
\newcommand{\reqBlocking}{R3}
\newcommand{\reqDeniableSend}{R4}
\newcommand{\reqDeniableReceive}{R5}
\newcommand{\reqDeniableBlocking}{R6}
\newcommand{\reqPublicKey}{R7}

\else
\newcommand{\reqDeniableSend}{R1}
\newcommand{\reqDeniableReceive}{R2}
\newcommand{\reqDeniableBlocking}{R3}
\newcommand{\reqPublicKey}{R4}
\fi

\newcommand\paragraf[1]{\noindent\textbf{#1}}

\newif\iftight
\tighttrue

\iftight
\let\OLDthebibliography\thebibliography
\renewcommand\thebibliography[1]{
  \OLDthebibliography{#1}
  \setlength{\parskip}{-0.1em}
  \setlength{\itemsep}{-0.2em}
}
\fi

\begin{document}
 
\title{\huge With a Little Help from My Friends: Transport Deniability for Instant Messaging}
\titlerunning{With a Little Help from My Friends}


\author{Boel Nelson \and
Aslan Askarov}
\authorrunning{B. Nelson \and A. Askarov}
%
\institute{Aarhus University \\
\email{\{boel,aslan\}@cs.au.dk}}

\maketitle

\begin{abstract}{
Traffic analysis for instant messaging (IM) applications continues to pose an important privacy challenge. In particular, transport-level data can leak unintentional information about IM -- such as who communicates with whom. Existing tools for metadata privacy have adoption obstacles, including the risks of being scrutinized for having a particular app installed, and performance overheads incompatible with mobile devices.
We posit that resilience to traffic analysis must be directly supported by major IM services themselves, and must be done in a low-cost manner without breaking existing features. As a first step in this direction, we propose a \emph{hybrid messaging model} that combines regular and \emph{deniable} messages. We present a novel protocol for \textit{deniable instant messaging}, which we call \ourprotocol. \ourprotocol\ is built on the principle that deniable messages can be made indistinguishable from regular messages with a little help from a user's friends. Deniable messages' network traffic can then be explained by a plausible cover story. \ourprotocol\ achieves overhead proportional to the messages sent, as opposed to scaling with time or number of users. To show the effectiveness of \ourprotocol, we implement a trace simulator, and show that \ourprotocol's deniability guarantees hold against strong \attackers\ such as internet service providers.
}\end{abstract}

  \keywords{instant messaging, plausible deniability, transport privacy}

\section{Introduction}
\label{sec:Introduction}
Instant messaging (IM) apps are popular, with seven billion registered accounts worldwide in 2019~\cite{the_radicati_group_inc_instant_2019}. Today, many IM services provide end-to-end encryption, with for example Signal, WhatsApp~\cite{whatsapp_whatsapp_2020}, Wire~\cite{wire_swiss_gmbh_wire_2021} and Facebook Messenger~\cite{facebook_newsroom_messenger_2016}, all using the formally secure~\cite{cohn-gordon_formal_2020} Signal protocol.

However, protecting metadata privacy of IM users is difficult. Traffic analysis remains an effective censorship mechanism~\cite{fu_service_2016,taylor_robust_2018}. 
Governments, organizations, and internet service providers (ISPs) filter websites in at least 103 countries~\cite{raman_measuring_2020}. China's ``great firewall'' actively probes and censors privacy tools~\cite{ensafi_examining_2015}. And, notably, critical decisions are based on metadata -- ``we kill people based on metadata'', as former US government official general Hayden~\cite{johns_hopkins_university_johns_2014} put it.

Modern IM services provide unique challenges and opportunities for metadata privacy. First, a large user-base is inherently censorship resilient since too blunt of a censorship can unite the public~\cite{roberts_resilience_2020} and cause a backlash. Despite IM having a user-base of 2.9B~\cite{jugovic_spajic_text_2019}, there exists no app with metadata privacy for IM. Even the Electronic Frontier Foundation (EFF) hesitates to recommend which IM to use from a privacy perspective~\cite{gebhart_why_2018}. Second, there is social pressure among individuals that push them to use the app their friends use, instead of the most secure app~\cite{gerber_why_2019}. Third, building an IM service with metadata privacy is nontrivial. The Tor messenger was a dedicated project for metadata private IM. Despite their efforts, Tor messenger failed to achieve satisfactory metadata privacy, ran out of funds, and ultimately had to be discontinued~\cite{the_tor_project_sunsetting_2018}. 
%


While the general problem of metadata privacy has been extensively studied, there are both social and technical barriers that prevent adoption of existing privacy tools to IM services.
%
On a social level, people are either unaware of privacy tools~\cite{ruogu_data_goes_everywhere_2015} or have diverse misconceptions about them~\cite{story_awareness_2021}. They, furthermore, find these tools too complicated or lack the knowledge how to use them~\cite{gerber_why_2019}. Beyond the challenges of usability, there is a risk of being put under scrutiny for having a particular app installed~\cite{samuel_china_2019,wagstaff_failing_2013}.

On a technical level, available tools are far from perfect. The Tor project~\cite{tor_project_tor_nodate} although popular with 2M active users~\cite{the_tor_project_users_2021}, is vulnerable to de-anonymization~\cite{karunanayake_-anonymisation_2021}, denial of service (DoS)~\cite{jansen_point_2019}, and traffic analysis~\cite{nasr_deepcorr_2018}. Because Tor can be automatically fingerprinted~\cite{fu_service_2016,taylor_robust_2018}, it is also easy to block (ironically, the authors of this paper themselves were blocked from accessing the Tor project's website on their organization network). Metadata private focused IM tools that run on Tor~\cite{zbay_how_nodate-1,briar_how_nodate,cwtch_overview_nodate,ricochet_refresh_nodate} suffer from the same issues. Other tools that hide traffic by imitating well-known apps do not produce credible traffic~\cite{houmansadr_parrot_2013}. 

The strongest guarantees for metadata privacy are provided by dedicated protocols. In particular, round-based protocols~\cite{wolinsky_dissent_2012,corrigan-gibbs_dissent_2010,van_den_hooff_vuvuzela_2015,lazar_karaoke_2018}, where predetermined rounds make traffic patterns indistinguishable, are able to resist traffic analysis. Unfortunately, round-based protocols are both resource exhaustive and inflexible. The rounds themselves require constant overhead, which results in poor performance~\cite{gilad_metadata-private_2019}, making them especially infeasible for resource constrained devices such as phones. Moreover, a major obstacle with round-based protocols is that they depend on fixed sets of individuals participating. That is, participants cannot join or leave without changing the privacy guarantees. Finally, round-based protocols are also easy to fingerprint and block.

The situation calls for a pragmatic approach. As EFF put it, ``An app with great security features is worthless if none of your friends and contacts use it''~\cite{electronic_frontier_foundation_communicating_2020}. Metadata private IM therefore needs to be incorporated into existing services to reach the masses. With 2.52B individuals using IM on phones~\cite{jugovic_spajic_text_2019}, deploying round-based protocols would be inconceivable. 

As a way forward, we propose a \textit{hybrid model} where regular and deniable traffic is combined in one protocol to create a hide-in-the-crowd effect. That is, our goal is to provide metadata privacy for \textit{some} traffic. While not all messages are deniable, all users have the option to send deniable messages. Notably, the hybrid model also aligns with the \textit{cute cat theory of censorship}~\cite{zuckerman_cute_2015} that posits that platforms that combine entertainment with political activism are more resilient to censorship than dedicated political platforms. Deploying a hybrid model therefore allows the app itself to stay innocuous -- it becomes both harder to block, and avoids putting individuals under scrutiny by its mere presence.

Another pragmatic aspect of our approach is centralization. All popular IM services are centralized~\cite{bahramali_practical_2020}. While far from ideal, when servers reside outside of censoring regimes, centralized models still protects against strong global adversaries. At the same time, centralized servers provide several design opportunities. First, server-side techniques can be both safer and easier to deploy than client-side techniques~\cite{bock_come_2020}. And secondly, centralization allows us to design protocols that are more scalable than zero trust protocols that scale with the number of users~\cite{gilad_metadata-private_2019}. 

In this paper, we present \ourprotocollong. \ourprotocol\ is a protocol designed to provide both message confidentiality and transport privacy for IM. Essentially, the idea behind \ourprotocol\ is to preserve the current functionality of IM services, and extend them with the option of sending deniable messages.

The design of \ourprotocol\ prioritizes the incentive to adoption by IM services, but as we explain below, comes at a price of additional latency for deniable messages. The protocol is based on the following principles. 

\begin{enumerate}[noitemsep,nolistsep]
    \item \emph{Hybrid messaging model}. Unlike cover protocols~\cite{houmansadr_i_2013,mohajeri_moghaddam_skypemorph_2012,sharma_camoufler_2021,wang_censorspoofer_2012}, we assume that the IM service is used for both regular, i.e., identifiable, and deniable communication. Users select which messages need to be communicated in a deniable manner. The hybrid model provides plausible deniability. For example, a whistleblower communicating with a journalist over \ourprotocol\ has a plausible explanation, \textit{a cover story}, supported by the network traffic -- they were talking to their friend, not the journalist.
    
    \item \emph{Privacy guarantees are independent of number of users}. Unlike round-based protocols, \ourprotocol\ does not provide privacy within anonymity sets, and privacy guarantees do not change over time.
    
    \item \emph{Lightweight protocol through store-and-forward trusted server.}  \ourprotocol\ relies on a simple centralized architecture where the messages are routed through a trusted server. Regular messages are delivered immediately. When forwarding the regular message, the server attaches a predetermined number of \emph{piggybacking} messages. Deniable messages are buffered on the server until they can be piggybacked with another regular message to the receiver.
    
    \item \emph{Asymmetric latency between regular and deniable communication}. When using \ourprotocol\ for regular communication, messages are forwarded immediately resulting in low latency overhead. For deniable communication, the latency depends on when messages can be safely piggybacked. We assume that a higher latency overhead is acceptable for deniable communication.
    
    \item \emph{Trusted contacts for decoy traffic}. Sending deniable messages requires users to select trusted contacts for receiving decoy messages from the server.
    
\end{enumerate}

The contributions of this paper are as follows.
\begin{itemize}[noitemsep,nolistsep]
    \item It proposes the hybrid messaging model for achieving deniability in IM services with low-barrier to adoption (\Cref{sec:setting}).
    \item It presents the core protocol \ourprotocollong\ (\Cref{sec:coreprotocol}), as well as its 
    practical extensions that avoid information leaks via key lookups (\Cref{sec:caching}) 
    \item It presents a privacy analysis (\Cref{sec:analysis}) and overhead evaluation (\Cref{sec:performance}) of the above protocols .
    \item It presents an enforcement mechanism, through a small programming language, to simulate realistic conversations given an \textit{interaction recipe} (\Cref{sec:interaction-recipes}). 
\end{itemize}

\section{Background on IM services}\label{sec:background}
In 2019, instant messaging (IM) services had seven billion registered accounts worldwide~\cite{the_radicati_group_inc_instant_2019}. The most popular IM services include WhatsApp (2B users), Facebook messenger (1.3B users), iMessage (estimated to 1B users), Telegram (550M users), and Snapchat (538M users)~\cite{statista_most_2021,kastrenakes_apple_2021}. While IM may appear deceptively simple, the sheer amount of users and traffic (41M messages/min~\cite{jugovic_spajic_text_2019}) present several engineering challenges. Keeping up with the demands, requires deploying and maintaining robust systems. As an example, WhatsApp's architecture handles over 1M connections per server~\cite{ixsystems_inc_rick_2014}, and maintaining the service occupies around 50 engineers~\cite{metz_why_2015}.

\ifverbose
\else
All major IM services, including WhatsApp, Facebook messenger, Telegram and Snapchat, use centralized servers to forward messages~\cite{bahramali_practical_2020}. Many IM apps also come with end-to-end encryption in addition to server-client encryption. Telegram uses their own protocol, MTProto~\cite{telegram_mtproto_nodate}, iMessage uses \rsaencryption~\cite{apple_inc_how_2021}, and Snapchat uses an unnamed encryption scheme for some of its content~\cite{salim_finally_2019}. WhatsApp~\cite{whatsapp_whatsapp_2020} and Facebook Messenger~\cite{facebook_newsroom_messenger_2016} on the other hand both use the Signal protocol~\cite{marlinspike_advanced_2013,marlinspike_x3dh_2016}. 
Of these protocols, Signal is the most popular, as it is also used by Wire~\cite{wire_swiss_gmbh_wire_2021}, ChatSecure, Conversations, Pond, the Signal app, and Silent Circle~\cite{cohn-gordon_formal_2020}. The Signal protocol is formally secure~\cite{cohn-gordon_formal_2020}, and is based on Off-the-Record Messaging (OTR)~\cite{borisov_off--record_2004} and Silent Circle Instant Messaging Protocol (SCIMP)~\cite{moscaritolo_silent_2012}. Despite strong security through encryption, none of the centralized IM services support deniable messaging.
\fi
\section{Threat model}\label{sec:setting}
\ifverbose
\else
In this section, we define our threat model, including \attacker\ capabilities, our notion of deniability, and which assumptions \ourprotocol\ operates under. 
\fi

\subsection{\Attacker}\label{sec:attacker}

We consider an active global external adversary~\cite{diaz_towards_2003} 
-- the \attacker\ can monitor, drop, or introduce traffic in the entire network. To exemplify: an active global external \attacker\ could be an internet service provider (ISP).

We assume that the \attacker\ knows which users use \ourprotocol\ and is aware of the protocol's hybrid model that allows users to send deniable messages. As the \attacker\ is active, they can also send messages to the users of \ourprotocol. The primary objective of the \attacker\ is to identify the presence of deniable messages. 

\subsection{Deniability}\label{sec:deniability}
\ourprotocol\ aims to achieve \textit{plausible deniability} for the deniable messages in the hybrid messaging model. 
\ifverbose
\else
Users must be able to plausibly deny both having sent, and having received a deniable message.
\fi
Plausible deniability here means that there is an alternative explanation, a cover story, for the user's network traffic. Since both sending and receiving messages can be denied, a user can also plausibly deny having communicated with a specific user. 







\subsection{Assumptions and trust}\label{sec:assumptions}

\ourprotocol\  relies on standard cryptographic assumptions. 
Moreover, to provide deniability, \ourprotocol\ depends on users getting a little help from their friends. In particular, to send each \deniablemessage, the sender needs to choose a \textit{trusted contact}, which is crucial to providing deniability. Each \deniablemessage\ may use a different trusted contact. We make the following assumptions.

\begin{description}[noitemsep,nolistsep]
    \item[\asmTrustedServer] A server is either trusted to distribute public encryption keys, or trusted to ensure deniability when forwarding deniable messages. We assume the servers cannot be compromised by the \attacker.
    \item[\asmTrustedDecoy] A trusted contact cannot be malicious, or be compromised by the \attacker.
    \item[\asmTrustedRcp] The receiver of a \deniablemessage\ cannot be malicious, or be compromised by the \attacker. Otherwise, they can reveal the identity of the sender.
    \item[\asmDecoyNotRcp] The trusted contact used as a decoy cannot be the same user as the receiver of a deniable message. Using regular messages to hide deniable messages to the same user would defeat the purpose of deniability.
\end{description}

\section{Our protocol}\label{sec:protocol}
This section presents \ourprotocollong\ -- a simple centralized IM protocol with message deniability. Crucially, while \ourprotocol\ benefits from having \normalmessages\ to hide \deniablemessages\ among, as we will later show, \ourprotocol\ maintains message deniability even when all messages are deniable. To divide trust, a \ourprotocol\ server either forwards messages, or distributes keys. 

\Cref{sec:coreprotocol} presents the high-level idea of the protocol; the rest of this section explains the details. Then, in \Cref{sec:caching} we extend the protocol to include caching of keys to improve performance. 

\subsection{Core \ourprotocol}
\label{sec:coreprotocol}
\begin{figure}[tb]
    \centering
    \begin{subfigure}[t]{0.5\linewidth}
    \centering
    \includegraphics[width=1\linewidth]{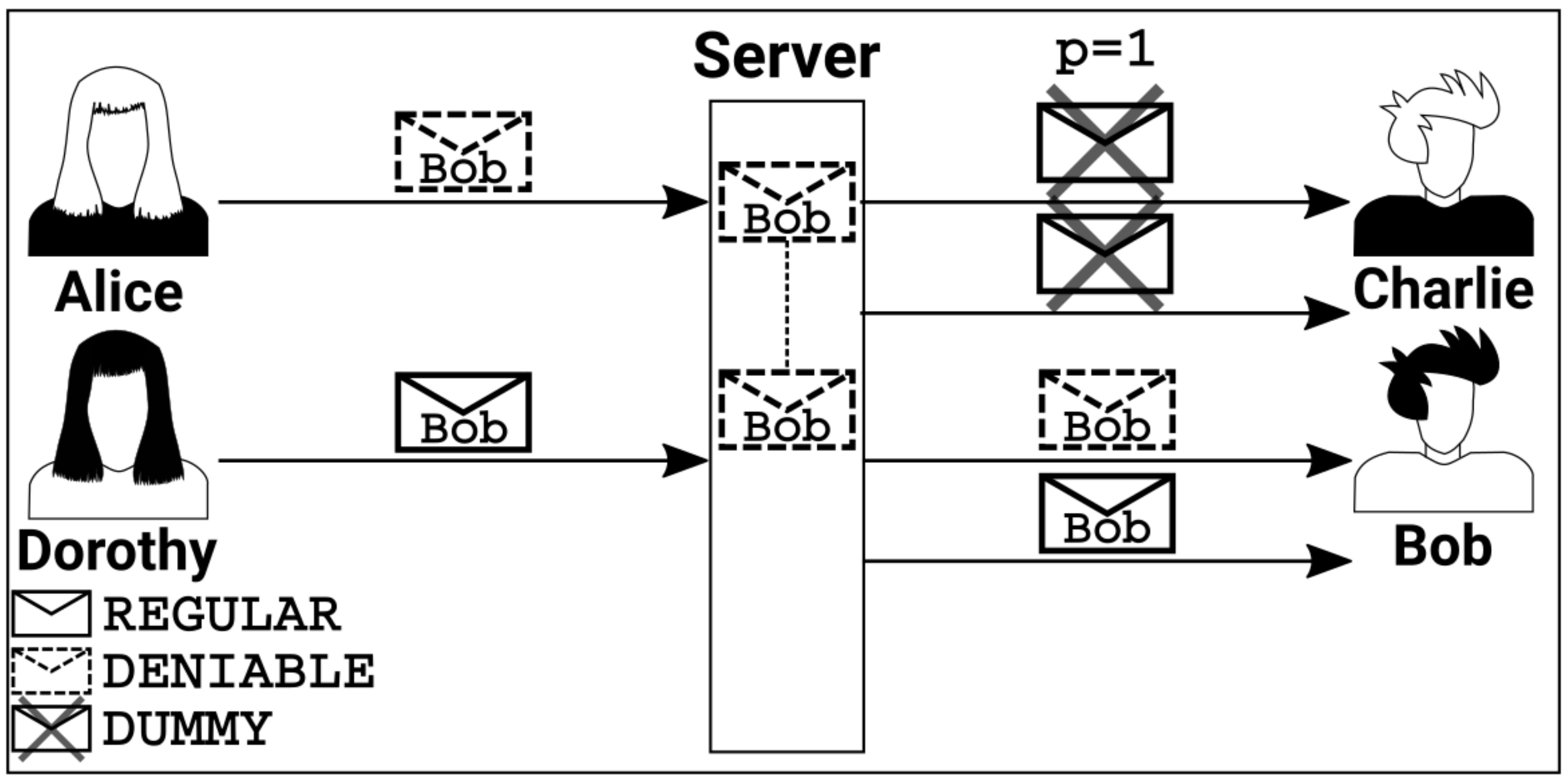}
    \caption{Alice sends a \deniablemessage\ to Bob using Charlie as her trusted contact. Alice's message piggybacks on Dorothy's \normalmessage\ to Bob.}
    \label{fig:protocol-example}
    \end{subfigure} \hfill
    \begin{subfigure}[t]{0.45\linewidth}
    \centering
    \includegraphics[width=1\linewidth]{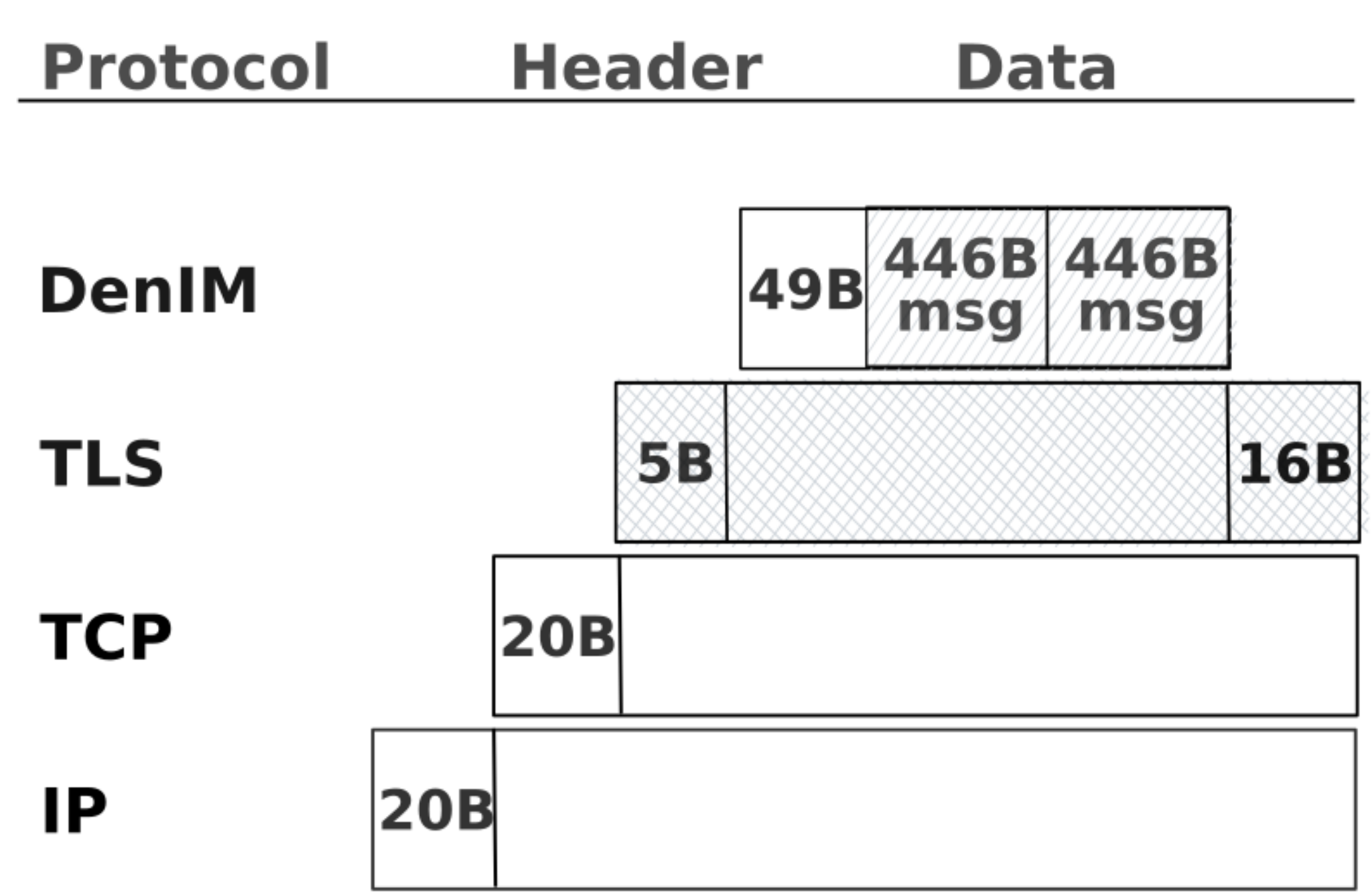}
    \caption{\ourprotocol\ data is encapsulated in an encrypted TLS data unit before it is passed down the network stack}
    \label{fig:denim-pdu}
    \end{subfigure}
    \caption{Protocol overview and protocol data unit for messages}\label{fig:denim}
\end{figure}

To exemplify, we start with a setting (\Cref{fig:protocol-example}) where each user's public key is pre-distributed, and \textit{Alice} wants to send a \deniablemessage\ to \textit{Bob}. First, Alice chooses a trusted contact, \textit{Charlie}, to use as a decoy that hides the presence of Alice's message. When Alice sends the \deniablemessage, the server immediately sends a \dummymessage\ to Charlie, and then queues Bob's deniable message. The \deniablemessage\ for Bob remains queued in the server until someone, e.g. \textit{Dorothy}, sends a \normalmessage\ to Bob. The server forwards Dorothy's message and adds $p$ piggybacking messages that contain Alice's \deniablemessage.

To protect against an \attacker\ (\Cref{sec:attacker}), \ourprotocol\ provides deniable versions of the following actions: key lookups, sending messages, forwarding messages, and blocking users from sending deniable messages.

\paragraf{Key lookup.} All users need to register their key before participating in \ourprotocol. The occurrence of key requests and responses are not hidden. Instead, we hide how many and which keys are requested. Key requests are encrypted and padded (see \Cref{app:data-formats}) to always be the size of two keys, as are key responses. 

\paragraf{Sending messages.}
\ourprotocol\ messages consists of headers that include meta-information -- sender, true receiver, decoy receiver, and message type (\flagnormal, \flagdeniable, \flagdrop, \flagblock) -- and a message payload. For confidentiality and to achieve constant size (\Cref{fig:denim-pdu}), we pad and encrypt (detailed description in \Cref{app:data-formats}) the payload with the receiver's key, and then add the headers and pad and encrypt again with the server's key. The resulting data fits within a common ethernet frame, and supports messages up to 892 bytes.


\Cref{code:sender-messages} presents the sender's code for sending messages.
The sender pads their message to a constant size and encrypts their message payload (line~\ref{line:regular-payload} and line~\ref{line:deniable-payload}, respectively), then creates a padded message object (line~\ref{line:regular-message} and line~\ref{line:deniable-message}, respectively), before forwarding it to the server. Senders are responsible for choosing a trusted contact (line~\ref{line:deniable-receiver}) for each \deniablemessage.

\begin{lstlisting}[language=TypeScript,label=code:sender-messages,caption={Sender's code to send \normalmessages, and \deniablemessages}]
sendRegularMessage(who: NetworkNode, what: string) { /*@ \label{line:regular-receiver} @*/
 let key = fetchKey(who)
 let payload = new EncryptedString(key, what) /*@ \label{line:regular-payload} @*/
 let msg = new PaddedMsg(this, who, null, /*@ \label{line:regular-message} @*/
               MessageType.REGULAR, payload)
 server.message(msg) }
sendDeniableMessage(decoy: NetworkNode, who: NetworkNode, /*@ \label{line:deniable-receiver} @*/
what: string) { 
 let keys = fetchKey(who, decoy)
 let payload = new EncryptedString(keys.1, what) /*@ \label{line:deniable-payload} @*/
 let msg = new PaddedMsg(this, who, decoy, /*@ \label{line:deniable-message} @*/
               MessageType.DENIABLE, payload)
 server.message(msg) }
\end{lstlisting}

When forwarding messages (\Cref{code:server-forwarding}), the server adds $p$ piggybacking messages. The server is responsible for padding all dummy messages to a constant size, in constant time. If there are \deniablemessage s in the queue, the server inserts them instead of dummy messages. We assume both these actions take the same time, which is practically achieved by timeboxing the functions. Incoming deniable messages are always queued \textit{after} the \normalmessage\ and the piggybacking messages have been forwarded to the receiver (line~\ref{line:queuing-deniable}).

\begin{lstlisting}[language=TypeScript,label=code:server-forwarding,caption={Server inserting piggybacking messages, and forwarding}]
forward(msg:PaddedMsg) {
 let receiver = msg.true_receiver
 for(let i = 0; i < p_value; i ++) { /*@ \label{line:loop} @*/
  let msg_p = null 
   if(deniable_queue.length > 0) {
    msg_p = deniable_queue.shift()
   } else {
	msg_p = new PaddedMsg(this, receiver, null,
                MessageType.DUMMY, new DummyString()) }
  receiver.message(msg_p) } //Deniable or dummy
 receiver.message(msg) } //Dummy or regular /*@ \label{line:forward-regular} @*/   
\end{lstlisting}

\begin{lstlisting}[language=TypeScript,label=code:server-receive,caption={Server receiving messages, notice how queuing always happens after messages have been forwarded}]
message(msg:PaddedMsg) {
 switch(msg.message_type) {
  case MessageType.REGULAR:
   this.forward(msg)
   break
  case MessageType.DENIABLE:
   this.forward(new PaddedMsg(this, msg.decoy_receiver,  /*@ \label{line:dummy} @*/ 
        null, MessageType.DUMMY, new DummyString()))
   if(!receiver.blocklst.get(msg.sender)) { //Blocked?/*@ \label{line:blocked} @*/
    this.deniable_queue.push(msg) } //Queue message /*@ \label{line:queuing-deniable} @*/
   break } }
\end{lstlisting}

\paragraf{Blocking deniable messages.} \ourprotocol\ supports receivers blocking senders from sending them \deniablemessages, without letting the sender find out they have been blocked. Our implementation of blocking (\Cref{code:sender-block}) essentially sends a deniable message addressed to the server. However, a block request uses the flag \flagblock\ instead of \flagdeniable. The server drops deniable messages sent by blocked users instead of queuing them (\Cref{code:server-receive}, line \ref{line:blocked}), but still sends a decoy message to the trusted contact. 

\begin{lstlisting}[language=TypeScript,label=code:sender-block,caption={A block request is a special case of a deniable message, where the intended receiver is the server}]
block(decoy: NetworkNode){
 let keys = fetchKey(decoy)
 let payload = new EncryptedString(decoy.1, "")
 let msg = new PaddedMsg(this, server, decoy, /*@ \label{line:server-receiver} @*/
               MessageType.BLOCK_REQUEST, payload)
 server.message(msg) }
\end{lstlisting}

\subsection{\ourprotocolCaching}\label{sec:caching}
To avoid requiring key lookups for \textit{each} message, we support caching of keys. \ourprotocol\ uses a partitioned cache, divided in a regular cache, and a cache for $\langle$deniable,decoy$\rangle$ key pairs. Each cache entry has a time to live (TTL) that invalidates the key.
The use of the caches is regulated by the following rules.
\begin{description}[noitemsep,nolistsep]
\item [To send a regular message:]\ 
\begin{enumerate}[label=R$_\arabic*$]
    \item If receiver is alive in regular cache: re-use key, bump TTL \label{case:normal-cache-reuse}
  \item If receiver is alive as decoy in deniable cache: re-use key, bump TTL for the pair's entry in deniable cache \label{case:decoy-cache-reuse}
  \item Otherwise: fetch receiver's key, store in regular cache \label{case:normal-no-reuse}
    \item Send message
\end{enumerate}
\item[To send a deniable message:] \ 
\begin{enumerate}[label=D$_\arabic*$]
    \item\label{case:decoy-in-normal-use} If decoy is alive in regular cache: abort execution
  \item\label{case:decoy-in-use} If decoy is alive and tied to another receiver in deniable cache: abort execution 
  \item\label{case:decoy-reuse} If decoy is alive and already tied to receiver: re-use key, bump TTL  
  \item[\label{case:deniable-fetch}] Otherwise: fetch decoy and receiver's key, store together in deniable cache
    \item Send message
\end{enumerate}
\end{description}
Note how cases \ref{case:decoy-in-normal-use} and \ref{case:decoy-in-use}
abort the execution if the decoy is cached. That is, the sender can either wait until the trusted contact's key is invalidated in cache, or choose a new trusted contact for the message at hand. We leave the choice of strategy to future research as it should be based on user experience, which is beyond the scope of this paper. 

\section{Privacy analysis}\label{sec:analysis}
In this section, we show that sending \normalmessages, and sending \deniablemessages\ are indistinguishable in \ourprotocol\ to our \attacker. We also show that the presence of \deniablemessages\ among \dummymessages\ being forwarded by the server is unobservable to the \attacker. Since both sending and receiving \deniablemessages\ are hidden, the \attacker\ does not learn who communicates deniably with whom. Our use of a hybrid message model does however allow the \attacker\ to learn some information from observing the system. When the \attacker\ observers all $n$ messages they will learn that: \textit{at most} $n$ \deniablemessages\ were sent, silent users did not send any \deniablemessages, and users who did not receive any traffic did not receive any \deniablemessages.

Due to the store-and-forward design of our server, sending and receiving of messages are effectively disconnected. The server never directly forwards the deniable message it receives (\Cref{code:server-receive}, line~\ref{line:dummy}), which allows us to analyze send and forward as independent events. To support our analysis, we implement a trace-based simulator\footnote{Source code available at: \url{https://www.dropbox.com/sh/5mv8n2yub0gobwv/AAAh6LOXZ\_NZ1oZbaRu4xCQMa?dl=0}} of \ourprotocolCaching. Our implementation allows us to run arbitrary scenarios, and evaluate the network traces they generate. Crucially, \ourprotocol's privacy guarantees rely on the following principles:
\begin{description}[noitemsep,nolistsep]
    \item[Transport level indistinguishability] via separately encrypted headers and payloads, as well as padding to achieve TCP segments of constant size.
    \item[Plausible deniability for communication patterns] by ensuring the network trace generated by sending a \deniablemessage\ via a trusted contact, $C$, is consistent with a cover story consisting of sending a \normalmessage\ to $C$.
\end{description}
\noindent To simplify our analysis, we introduce a new assumption: 
\begin{description}[noitemsep,nolistsep]
    \item[\asmPatternContainment] Regular communication patterns between the sender and the trusted contacts \textit{contains} the patterns used by the sender for deniable messages.
\end{description}
Assumption \asmPatternContainment\ means that deniable messages do not noticeably influence the behavior of the users of \ourprotocol. Hence, we can view the sender and receiver as independent parties when reasoning about deniable traffic. Additionally, assumption \asmPatternContainment\ allows us to ignore the timing aspects of when and how many messages are sent. 
In \Cref{sec:interaction-recipes} we provide a mechanism to enforce assumption \asmPatternContainment.

\subsection{Indistinguishable data formats}\label{sec:analysis-formats}
The \attacker\ wants to distinguish between a user's true behavior and their cover story by observing the data transmitted over the network. Two TCP segments are indistinguishable to an attacker when: the segments have the same size, and the encrypted data appears indistinguishable from random strings to the \attacker, i.e. the encryption scheme provides indistinguishability under chosen cipher text attacks (IND-CCA). 

\paragraf{Key lookup.}
\ourprotocol\ does not try to hide the presence of key lookups. Instead, the amount of keys requested, and delivered, is secret as the number of keys would otherwise leak whether a trusted contact is needed or not. Our data format for key responses and key requests are encrypted providing IND-CCA, and padded to constant size (see \Cref{app:data-formats}). Accordingly, key requests/responses with one and two keys are indistinguishable on transport level. 

\paragraf{Messages.}
\Normalmessages, \deniablemessages\ and \dummymessages\ are encrypted to provide IND-CCA, have constant size headers, and are padded to have the same length (see \Cref{app:data-formats}). Hence, all messages in \ourprotocol\ are indistinguishable from each other on transport level.

\subsection{Indistinguishable mix of traffic}\label{sec:analysis-actions}
The \attacker\ can either passively observe, or actively try to influence the users' communication patterns, to try to deduce information about \deniablemessages.

\paragraf{Sending deniable messages.} The \attacker\ wants to distinguish between a user's true behavior and their cover story by observing the traffic between a sender and server. We will show that traffic generated by sending a deniable message is indistinguishable from its corresponding cover story, both with and without requiring a key lookup.

\textit{Without key lookup.}
Given a receiver $B$, and some trusted contact, $C$, sending a \deniablemessage\ to $B$ via $C$ must be indistinguishable from sending a \normalmessage\ to $C$. For the \deniablemessage: the server immediately forwards a \dummymessage\ to $C$, then queues the \deniablemessage\ for $B$. For the \normalmessage: the server immediately forwards the message to $C$. Since queuing is the only difference between the two actions, and queuing happens after any network traffic, the actions are indistinguishable to the \attacker.

\textit{With key lookup.} Given a receiver $B$, and some trusted contact, $C$, and a deniable cache without keys $\langle B_k,C_k\rangle$, key lookups triggered by sending a \normalmessage\ to $C$ must be indistinguishable from the key lookups triggered by sending a \deniablemessage\ to $B$ via $C$. The subtlety of caching keys lies in realizing which keys may be re-used, and that re-use is necessary.

From our algorithm in \Cref{sec:caching}, it follows that \ourprotocol\ enforces indistinguishability by: preventing (step \ref{case:decoy-in-normal-use} and \ref{case:decoy-in-use}) sending \deniablemessages\ using $C$ if $C_k$ is already in the regular cache (such a lookup leaks that $C$ is not the intended receiver), and only looking up $\langle B_k,C_k\rangle$ when $C_k$ is not in the regular cache (step \ref{case:decoy-reuse}), which is indistinguishable from looking up only $C_k$ by design. $\langle B_k,C_k\rangle$ is then treated and re-used as if $C_k$ is present in the regular cache. Hence, $C_k$ in regular cache behaves the same way $\langle B_k,C_k\rangle$ behaves in the deniable cache, which makes their corresponding re-use and timeout indistinguishable.




\paragraf{Forwarding deniable messages.} The \attacker\ wants to distinguish between \deniablemessages\ and \dummymessages\ by observing the timing of the server's added piggybacking traffic. We show that the presence of \deniablemessages\ among the piggybacking messages is unobservable.

Piggybacking traffic always consists of $p$ messages, which are indistinguishable from each other (\Cref{sec:analysis-formats}). The timing when sending the messages does not depend on if there are any deniable messages queued or not, since the server always loops on $p$ (\Cref{code:server-forwarding}, line~\ref{line:loop}) and we assume popping from the queue and creating a new message is timeboxed to execute in the same time. Accordingly, piggybacking traffic is always created in constant time, with TCP segments appearing random to the \attacker, making \deniablemessages\ unobservable.

\paragraf{Preventing active attacks.}
Due to assumption \asmTrustedRcp, users will not reply to an \attacker's \deniablemessages. The \attacker\ can still try to send \deniablemessages\ to users to provoke reactions. An \attacker\ dropping packets in \ourprotocol\ only affects quality of service, not privacy.

\textit{Deniable blocking.} An \attacker\ may try to exploit deniable blocking, by filling a user's deniable buffer in the server to provoke the user to eventually block them. Observing that they got blocked would leak that, with high probability, the \attacker's \deniablemessages\ have been delivered. As a defense, \ourprotocol\ supports users blocking other users from sending them deniable messages.

Deniable blocking is a special case of sending a deniable message: it is a deniable message addressed to the server. The only difference is that instead of queuing the message, the message is silently dropped. Since queuing happens after forwarding the decoy message to the trusted contact (\Cref{code:server-receive}, \ref{line:blocked}), the behavior is indistinguishable from the standard deniable messaging behavior. It follows that blocking is indistinguishable from other messages to an \attacker.

\textit{No delivery receipts for deniable messages.}
An \attacker\ could try to infer the status of the deniable buffer from delivery receipts of messages. By sending deniable messages to a user, if the \attacker\ got a delivery receipt, it would imply the buffer has been emptied (or was empty). To avoid leaking information about the deniable message buffers, we do not support delivery receipts for deniable messages.

\section{Enforcing assumption \asmPatternContainment\ via interaction recipes}\label{sec:interaction-recipes}
To enforce that communication patterns between the sender and the trusted contacts are realistic, when trusted contacts receive dummy message, we introduce the notion of \emph{interaction recipes}. 
Interaction recipes are small programs with instructions for when and how many messages to send back to the initiator. 
To program interaction recipes, we design a small programming language and a corresponding execution environment.
Interaction recipes are restricted by design, allowing nothing but basic control flow operations (branching and loops), integer built-in types, and calls to a few built-in functions. We use C-like surface syntax for writing interaction recipes and compile them to a simple JVM-like bytecode, for which we also have a special interpreter. Because interaction recipes execute on the devices of trusted users, a special minimal virtual machine reduces the complexity and the trusted computing base of \ourprotocol\ infrastructure. A trusted user's device only executes interaction recipes from their friends. 
Once compiled, interaction recipes are delivered as part of a \deniablemessage, using one of the 446B chunks (\Cref{fig:denim-pdu}) dedicated for message content.

We note that simulating convincing traffic patterns is a challenging~\cite{geddes_cover_2013,houmansadr_parrot_2013} research area, that requires complicated models of user behavior, such as OUStral~\cite{lorimer_oustralopithecus_2021}, to successfully avoid detection. In this paper we do not attempt to model user behavior, but propose and implement a mechanism (interaction recipes) that helps ensure that users conform to a desired communication pattern. 
\paragraph{Execution model and built-ins.}
Interaction recipes are run once they are received on the trusted contact's device; they can only communicate 
with the user that supplies them. The execution environment contains a number of useful built-ins. Built-in \texttt{wait} blocks until a particular event, such as app activation, is triggered, while built-ins \texttt{sleep}, and \texttt{usleep} implement sleeping functionality (in seconds and milliseconds, respectively).  Built-ins \texttt{store} and \texttt{load} implement persistent storage between different interaction recipes, confined to the same user; the storage is indexed through integer registers. Built-in \texttt{reset} terminates execution of the user's prior still-running interaction recipes, keeping the persistent storage.
\paragraph{Examples.}
To illustrate expressiveness of interaction recipes, we show three examples, that once compiled all fit well within 446B.  The examples simulate the following behaviors of the trusted contact: (i) responding when app is active (\Cref{code:recipe-example-app-active}), (ii) initiating a conversation at a given time (\Cref{code:recipe-example-reply-midnight}), and (iii) responding once to each message, but after some delay (\Cref{code:recipe-example-proportional-to-received}). For reader's convenience, the listings below italicize the built-ins.

\begin{lstlisting}[language=Noiselet,label=code:recipe-example-app-active,caption={A random number of messages is sent when the IM app is active, and the keyboard has not been used recently. Compiles to 31B.}]
int loop = 1;
while (loop) {
  wait(APP_ACTIVE); /* Wait until the app is active */
  if(gettime()-last_kb_time()>5){ /*Keyboard unused */
    send(rnd(1,4)); /* Reply back with 1 to 4 messages */
    loop = 0; }}
\end{lstlisting}

\begin{lstlisting}[language=Noiselet,label=code:recipe-example-reply-midnight,caption={A conversation is initiated, here at midnight. Compiles to 23B.}]
int today_time = gettime()%86400; /*Days: 60x60x24*/
int to_wait = 86400 - today_time;
sleep(to_wait); /*Clock is 00:00*/
send(1);
\end{lstlisting}

\begin{lstlisting}[language=Noiselet,label=code:recipe-example-proportional-to-received,caption={A response is sent to each message received, with a delay. Compiles to 71B.}]
int last_msg = load (100); /*Read state of last message*/
int msgs = 1; /*Number of messages received*/
reset(); /* Kill other active recipes */
store(100, gettime()); /* Store timestamp for message */
if(load(0)) { /* No previous messages received */
  store(0,msgs); /* Initialize count */ } else {
  msgs = load(0)+1; /* Increment previous count */
  store(0,msgs); } /* Store updated value */
sleep(30); /* Wait for potential interruption */
while(msgs) { /* For each message received... */
  usleep(rnd(1000,5000)); /* Wait between 1-5 seconds */ 
  send(1); 
  msgs = msgs - 1;
  store(0,msgs); } /* Update message count */
\end{lstlisting}
\section{Overhead and quality of service analysis}\label{sec:performance}
This section presents an overhead analysis of \ourprotocol. The \emph{anonymity trilemma}~\cite{das_anonymity_2018} states that a protocol for anonymous communication can only achieve two of the following simultaneously: strong anonymity, low bandwidth overhead and low latency. Since \ourprotocol\ achieves strong anonymity through deterministic guarantees, \ourprotocol\ is forced to introduce either bandwidth or latency overhead. For comparison, we assume a baseline IM protocol without deniability that uses the same key length (512B) and encryption scheme (\rsaencryption\ for message content, and \tlscipher\ for headers and key lookups) as \ourprotocol. We assume that in the baseline protocol, all messages are regular, and headers consists of two 16B UUIDs to identify sender and receiver. 

\subsection{Bandwidth and latency overhead}
We analyze overhead introduced by each kind of the traffic.

\paragraf{Key lookup.} Bandwidth overhead of key requests and responses are doubled with respect to message content in \ourprotocol\ compared to the baseline. Latency is primarily added by symmetric encryption through TLS, which is performed once per response/request both in \ourprotocol\ and the baseline, therefore we assume the difference is negligible. We also assume the latency overhead from two keys as opposed to one key is negligible, as the complexity of each lookup is $\mathcal{O}(1)$. Padding adds no latency, as we can safely repeat the requested key twice in the plain text due to using an IND-CCA encryption scheme, without giving the \attacker\ an advantage.

\paragraf{Sending messages.} With \rsaencryption\ encryption using 512B keys, and TLS used for headers, the smallest baseline message is 605B (2*16B + 512B + 5B + 16B + 20B + 20B), and can carry text of up to 446B. A \ourprotocol\ message is always \pdusize, and carries text of size up to 892B, text larger than 892B will be divided into multiple messages. For messages, $m_i \leq$ 446B, the bandwidth overhead per message is 529B, and for messages 446B $< m_j \leq$ 892B, the overhead is 17B.

We assume the latency difference between encrypting the baseline's headers (32B) and \ourprotocol's headers (49B) with TLS is negligible since the encryption is symmetric. To encrypt the messages with \rsaencryption, we divide the message into 446B chunks and encrypt each chunk in constant time. For messages $m_i \leq$ 446B, \ourprotocol\ is $\mathcal{O}(1)$ slower than the baseline, and for messages 446B $< m_j \leq$ 892B, the latency is the same.

\paragraf{Forwarding messages.} For each regular message forwarded, \ourprotocol\ adds $p$ \pdusize\ messages. We will interpret dummy messages as overhead, and not count deniable message as overhead. Given a deniable buffer with $n$ deniable messages, the bandwidth overhead is $p-n$ * \pdusize. The latency overhead added is proportional to $p$.

\subsection{Quality of service}
The liveness of deniable messages is affected by how much help users get from their friends. By design, users that do not receive any regular traffic, also do not receive deniable traffic. Since dummy messages used for decoy traffic are treated as regular messages, being utilized as a trusted contact helps avoid starvation. A pragmatic solution is to implement periodic polling to prevent starvation.

\section{Discussion and future work}\label{sec:discussion}
This section discusses design aspects of \ourprotocol, and future directions for the protocol.

\paragraf{Centralized architecture.}
Centralization in \ourprotocol\ is a pragmatic design choice. Our position is that a lightweight centralized protocol is more likely to be adopted by a major service than a federated alternative.  Any form of deniability is better than none. There is evidence to support this position. Case in point is WhatsApp's use of Noise Pipes that reduces the amount of exposed metadata~\cite{perrin_noise_2018,whatsapp_whatsapp_2020}. We speculate that access to deniability in a major service, however constrained, will cultivate privacy awareness among the user base, and lay ground for stronger mechanisms in the future.

Having said that, despite perceived simplicity, even in a centralized setting, metadata privacy is tricky. As we show in \Cref{sec:protocol} there are subtleties with caching keys, blocking, message size padding, and constant response time must be accounted.



\paragraf{Trust.} \ourprotocol\ assumes that receivers are trusted, and hence are allowed to know who communicates with them deniably. Nonetheless, assuming that receivers are trusted does not stop them from revealing the sender's secrets in other ways. For example, a receiver could reply with regular messages to a deniable message. We expect that this potential leakage could be solved in software. Future work could explore whether preventing the user from responding with regular messages, or nudging the user by asking if they may want to make their reply deniably, would be a better solution.

While we assume servers are trusted (assumption \asmTrustedServer), \attackers\ can still view servers as potential attack targets. To minimize the information held by forwarding servers at any point in time, real deployments should use small deniable buffers, which can be combined with timeouts for deniable messages. Still, too small buffers or too short timeout time could affect quality of service and result in undelivered deniable messages. A potential solution would be to jointly tune message timeout and periodic polling by clients.

\paragraf{Assumptions on trusted contacts.}
Our privacy analysis relies on our assumption \asmPatternContainment, which we supply the interaction recipe mechanism to enforce. Identifying what constitutes a 'good' interaction recipe is still a user dependent, unsolved problem.
As future work, we envision client software helping to identify when it is safe to send deniable messages based on users' past activity, and which of the user's trusted contacts is appropriate to use in a given time window. 

\section{Related work}\label{sec:related-work}
Unger~et~al.~\cite{unger_sok_2015} divide secure messaging into three different key challenges: \textit{trust establishment}, \textit{conversation security}, and \textit{transport privacy}. With \ourprotocol, we address transport privacy. We compare \ourprotocol\ to works that also target transport privacy, where \ourprotocol\ stands out by being the only one using a hybrid messaging model. Most similar to \ourprotocol\ is Camoufler~\cite{sharma_camoufler_2021}, which uses IM traffic to tunnel censored content. Since Camoufler is implemented on top of the Signal app, the network architecture is the same as ours, with trusted servers. 

\subsection{Transport privacy}
The perhaps most well-known implementation of transport privacy is the Tor browser~\cite{tor_project_tor_nodate,dingledine_tor_2004}, which uses onion routing~\cite{reed_anonymous_1998}. A key difference between our work and Tor is that we provide \textit{optional} deniability through our hybrid model for regular and deniable messages.

\paragraf{Cover-protocols: }
Some protocols like CensorSpoofer~\cite{wang_censorspoofer_2012}, SkyeMorph~\cite{mohajeri_moghaddam_skypemorph_2012}, and FreeWave~\cite{houmansadr_i_2013}  generate cover traffic to imitate another protocol. With \ourprotocol, the idea is instead that deniable messages are incorporated and delivered as part of the IM app, as opposed to generating traffic masquerading as IM traffic.

\paragraf{Round-based protocols: }
DC-nets~\cite{chaum_dining_1988} achieves privacy by making the presence or absence of a secret message indistinguishable by using communication rounds.

\ifverbose
\else
Dissent~\cite{corrigan-gibbs_dissent_2010} uses a version of DC-nets where users can initiate new communication rounds. In \ourprotocol, all communication is initiated by users. Anonycaster~\cite{head_anonycaster_2012} is a distributed DC-net where some nodes are assumed to be trusted by the user. In \ourprotocol, trusted contacts can be viewed as trusted nodes.

Several of the round-based solutions offer probabilistic privacy guarantees through differential privacy. These include Vuvuzela~\cite{van_den_hooff_vuvuzela_2015}, Stadium~\cite{tyagi_stadium_2017}, and Karaoke~\cite{lazar_karaoke_2018}, none of which use centralized servers like in \ourprotocol.

Connected to Vuvuzela is Alpenhorn~\cite{lazar_alpenhorn_2016}, which is implemented on top of Vuvuzela. Alpenhorn extends Vuvuzela to make key exchanges, which otherwise would happen outside of rounds, indistinguishable. \ourprotocol\ includes key lookups as part of the deniable traffic, which is possible due to the use of the hybrid messaging model.

Similar to Vuvuzela and Stadium, Atom~\cite{kwon_atom_2017} is focused on scalability for DC-nets. Since \ourprotocol\ does not use round-based communication, Atom's scalability solutions are not applicable to \ourprotocol.

Verdict~\cite{corrigan-gibbs_proactively_2013} is, like \ourprotocol, centralized. Still, Verdict's main contribution is prevention of jamming attacks in DC-nets, where users disrupt the communication rounds to attack availability. In comparison, jamming attacks are not applicable to \ourprotocol.
\fi

\paragraf{Broadcasts: }
Bitmessage~\cite{warren_bitmessage_2012} and Riposte~\cite{corrigan-gibbs_riposte_2015} broadcasts messages to groups of subscribers. Unlike \ourprotocol, they do not achieve sender deniability.

\paragraf{Delays: } 
Mixminion~\cite{danezis_mixminion_2003} introduces random delays to traffic. In contrasts, the delays in \ourprotocol\ are based on when \normalmessages\ can be piggybacked on.

Loopix~\cite{piotrowska_loopix_2017} is a mix-network~\cite{chaum_untraceable_1981} which uses layers of servers to forward messages. In addition to layers, Loopix also delays messages to make them unlinkable. With \ourprotocol\ there is no need for layers or delays, as the use of a hybrid messaging model allows us to create unlinkability by effectively hiding deniable messages completely.

IMProxy~\cite{bahramali_practical_2020} achieves privacy by inserting dummy traffic and introducing random delays at proxies in the network. While we do not add delays in \ourprotocol, we also insert dummy traffic. Still, the dummy traffic in \ourprotocol\ is deterministic, and the dummy traffic in IMProxy is probabilistic. Moreover, our threat models are different, as a centralized trusted server allows us to thwart global \attackers, while IMProxy only can protect traffic between proxies.

\paragraf{Private information retrieval (PIR): }
Pynchon~\cite{sassaman_pynchon_2005} and Pung~\cite{angel_unobservable_2016} achieve receiver deniability using the cryptographic primitive PIR. In comparison \ourprotocol\ also achieves sender deniability, but comes at the cost of trusted servers.

\subsection{Other versions of deniability}
\ifverbose
\else
A stronger notion of deniability than we use is online deniability~\cite{dodis_composability_2009}, adopted by for example GOTR~\cite{liu_improved_2013}. In online deniability one party in the deniable communication colludes with the \attacker. Since we assume senders and receivers trust each other, \ourprotocol\ does not offer online deniability.

Spawn~\cite{unger_deniable_2015} is a protocol for deniable authenticated key exchange (DAKE), which allows users to plausibly deny message transmission or having participated in a conversation. Similarly, \ourprotocol\ allows users to plausibly deny message transition, but not participation. Unlike \ourprotocol, Spawn's transmission deniability does not achieve transport privacy. Deniable encryption~\cite{canetti_deniable_1997} is also different from the type of deniability \ourprotocol\ offers. With deniable encryption, a user can plausibly deny that they can decrypt a specific cipher text.

Buddies~\cite{wolinsky_hang_2013} is an architecture that guarantees pseudonymity by making sets of users' traffic indistinguishable from each other. This is different from \ourprotocol, since we hide the presence of deniable messages completely.
\fi
\section{Conclusion}\label{sec:conclusion}

\ifverbose
\else
We design a novel protocol for \ourprotocollong, which utilizes regular IM traffic to hide deniable messages through adopting a \textit{hybrid message model}. By piggybacking deniable traffic on regular traffic, \ourprotocol's overhead scales proportional to the number of actions, as opposed to scaling with time or users. Crucially, the privacy of \ourprotocol\ does not depend on the amount of users in the system.  
\fi

\newpage
\bibliographystyle{splncs04}
\bibliography{references.bib}

\appendix

\section{Data formats}\label{app:data-formats}
\paragraf{Messages:}
We use TLS v1.3 to encrypt \ourprotocol\ headers, and RSAES-OAEP~\cite{jonsson_pkcs_2003} to encrypt payloads (messages). Assuming a key length of 4096b, which is a common key length for RSA, a 446B plain text will encrypt to 512B. As such, our encrypted payload will be a multiple of 512. To fit within an maximum transfer units (MTU), commonly 1500 bytes for Ethernet~\cite{hornig_standard_1984} and 2304 bytes for wifi~\cite{the_working_group_for_wlan_standards_ieee_2021}, we choose to support two encrypted plain texts (892B plain text which encrypts to 1024B).

After concatenating \ourprotocol\ headers (49B) and TLS headers (5B) to the encrypted message, the data is 1078B. The 1078B are then encrypted with TLS (assuming cipher suite \tlscipher), resulting in 1094B, and TCP and IP layers then adds 20B each for a total of \pdusize.


\paragraf{Key request:}
Similarly, key requests contain two UUIDs (one may be null), i.e. 32 byte. With TLS headers and using cipher suite \tlscipher, the total size is 53.

\paragraf{Key response:} We always send two keys (or padding) of size 512B each. The content and a TLS header becomes 1029 bytes. Encrypted with TLS (\tlscipher), the total size is 1045.

\end{document}


\end{document}